\documentclass[sigconf]{acmart}
\usepackage{geometry}
\usepackage{multirow}
\usepackage{graphicx}
\usepackage{syntax}
\usepackage[medium,compact]{titlesec}
\usepackage{marvosym}
\usepackage{enumitem}
\usepackage{tabularx}


\copyrightyear{2024}
\acmYear{2024}
\setcopyright{rightsretained}
\acmConference[DAC '24]{61st ACM/IEEE Design Automation Conference}{June 23--27, 2024}{San Francisco, CA, USA}
\acmBooktitle{61st ACM/IEEE Design Automation Conference (DAC '24), June 23--27, 2024, San Francisco, CA, USA}
\acmDOI{10.1145/3649329.3657356}
\acmISBN{979-8-4007-0601-1/24/06}

\begin{document}

\title[]{Data is all you need:  Finetuning LLMs for chip design via an automated design-data augmentation framework}

\begin{abstract}
Recent advances in large language models have demonstrated their potential for automated generation of hardware description language (HDL) code from high-level prompts. Researchers have utilized fine-tuning to enhance the ability of these large language models (LLMs) in the field of Chip Design. However, the lack of Verilog data hinders further improvement in the quality of Verilog generation by LLMs. Additionally, the absence of a Verilog and electronic design automation (EDA) script data augmentation framework significantly increases the time required to prepare the training dataset for LLM trainers. This paper proposes an automated design-data augmentation framework, which generates high-volume and high-quality natural language aligned with Verilog and EDA scripts. For Verilog generation, it translates Verilog files to an abstract syntax tree and then maps nodes to natural language with a predefined template. For Verilog repair, it uses predefined rules to generate the wrong verilog file and then pairs EDA Tool feedback with the right and wrong verilog file. For EDA Script generation, it uses existing LLM(GPT-3.5) to obtain the description of the Script. To evaluate the effectiveness of our data augmentation method, we finetune Llama2-13B and Llama2-7B models using the dataset generated by our augmentation framework. The results demonstrate a significant improvement in the Verilog generation tasks with LLMs. Moreover, the accuracy of Verilog generation surpasses that of the current state-of-the-art open-source Verilog generation model, increasing from 58.8\% to 70.6\% with the same benchmark. Our 13B model (ChipGPT-FT\footnote{https://github.com/aichipdesign/chipgptft}) has a pass rate improvement compared with GPT-3.5 in Verilog generation and outperforms in EDA script (\emph{i.e.,} SiliconCompiler) generation with only 200 EDA script data.
\end{abstract}

\author{\normalsize
\textbf{Kaiyan Chang\textsuperscript{1,3}, Kun Wang\textsuperscript{9,2}, Nan Yang\textsuperscript{2,3},{Ying Wang}\Letter\textsuperscript{2},Dantong Jin\textsuperscript{12}, 
Wenlong Zhu\textsuperscript{2,3}, Zhirong Chen\textsuperscript{6,10},Cangyuan Li\textsuperscript{2,3}, \\ Hao Yan\textsuperscript{7,5}, 
Yunhao Zhou\textsuperscript{7,11}, Zhuoliang Zhao\textsuperscript{7,8},  Yuan Cheng\textsuperscript{7,4},  Yudong Pan\textsuperscript{2,3}, Yiqi Liu\textsuperscript{2,3}, Mengdi Wang\textsuperscript{2},\\
Shengwen Liang\textsuperscript{1}, Yinhe Han\textsuperscript{2}, Huawei Li\textsuperscript{1,3}, Xiaowei Li\textsuperscript{1,3}}\\
\textit{SKLP, Institute of Computing Technology, Chinese Academy of Sciences, Beijing, China\textsuperscript{1}\\
CICS, Institute of Computing Technology, Chinese Academy of Sciences, Beijing, China\textsuperscript{2}\\
University of Chinese Academy of Sciences\textsuperscript{3}, 
Nanjing University\textsuperscript{4}, 
Shanghai University\textsuperscript{5}, 
Zhejiang University\textsuperscript{6}\\
Shanghai Innovation Center for Processor Technologies\textsuperscript{7}, 
FuDan University\textsuperscript{8}\\
Hangzhou Institute for Advanced Study, University of Chinese Academy of Sciences\textsuperscript{9}\\
University of Illinois at Urbana Champaign\textsuperscript{10}, 
Shanghai Jiao Tong University\textsuperscript{11}\\
Zhejiang Lab\textsuperscript{12}\\
changkaiyan@live.com, wangying2009@ict.ac.cn
}}
\authornote{Ying Wang is the corresponding author.}

\maketitle

\section{Introduction}


\begin{figure*}[htbp]
\centering
\begin{minipage}[t]{0.3\linewidth}
\centering
\includegraphics[width=\textwidth]{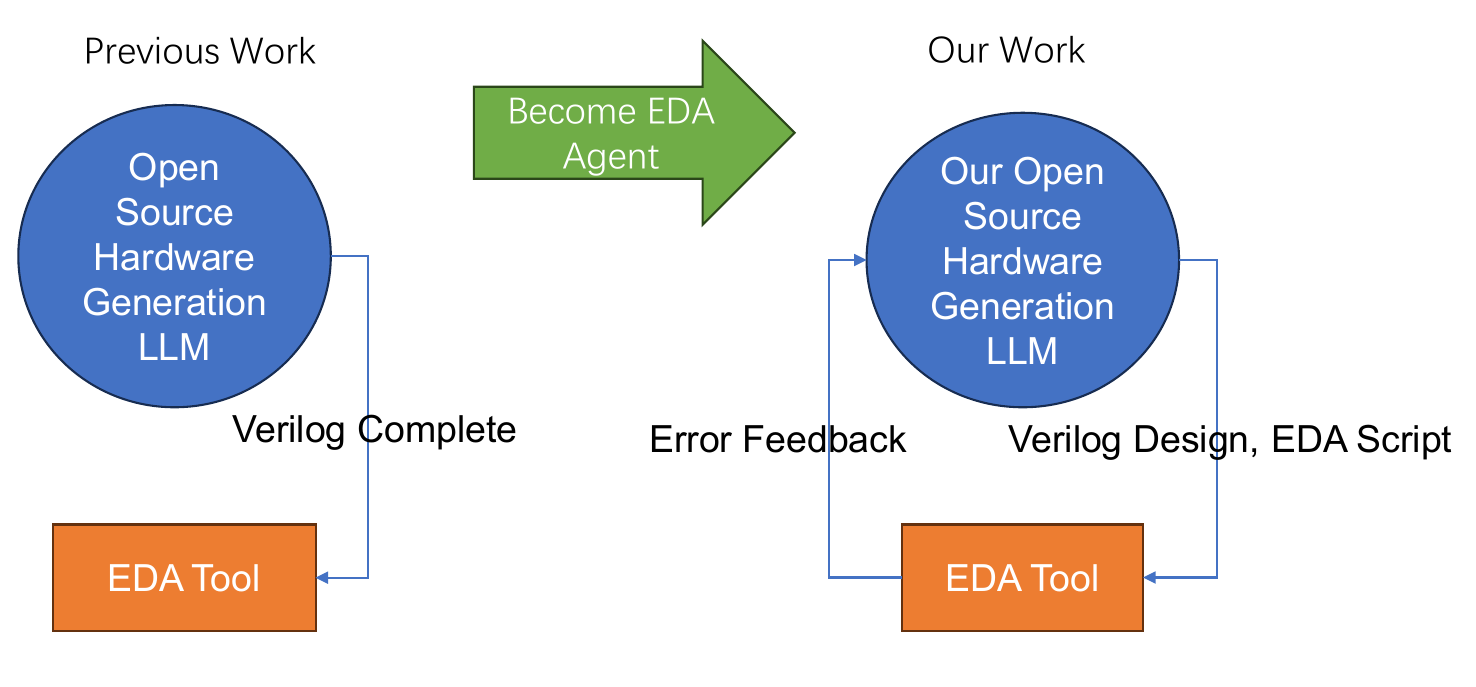}
\caption{Hardware Generation Large Language Model as an EDA Tool Agent.}
\label{fig:agent}
\end{minipage}
\begin{minipage}[t]{0.3\linewidth}
\centering
    \includegraphics[width=\textwidth]{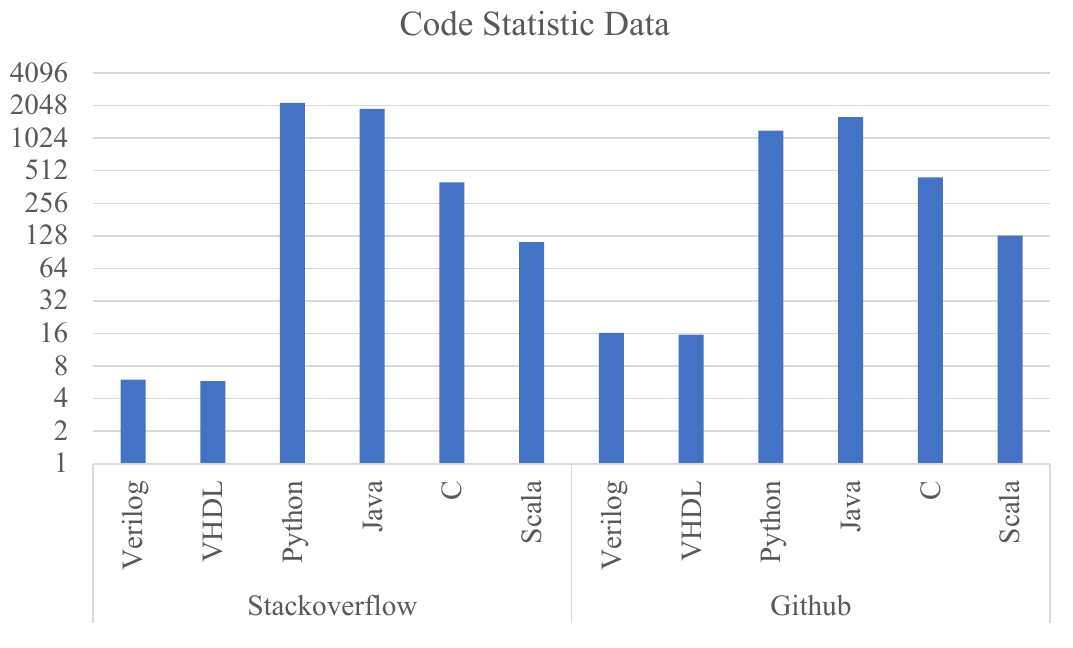}
    \caption{Compare different languages dataset scale. Hardware 
 language dataset is less than software language dataset.}
    \label{fig:codestatisticfile}
\end{minipage}
\begin{minipage}[t]{0.3\linewidth}
\centering
    \includegraphics[width=\linewidth]{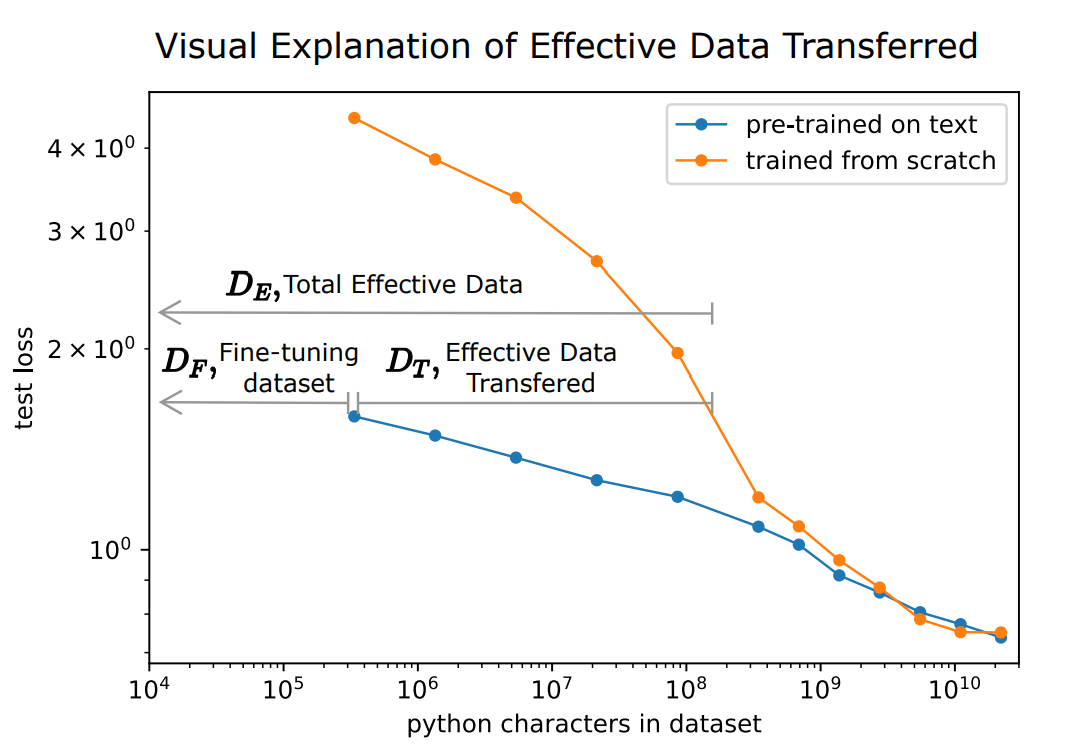}
    \caption{From OpenAI Technique Report for Scaling law in language model\cite{scalingtransfer}. As the dataset size improves, the loss decreases.}
    \label{fig:scalinglaw}
\end{minipage}
\end{figure*}
Recent advances in large language models (LLMs) have demonstrated great potential for automated generation of hardware description language (HDL) code from high-level prompts\cite{benchmarking,pearce2020dave,chang2023chipgpt} and EDA flow coordination \cite{he2023chateda}. It sheds light on the approach to agile chip development through natural language hardware design, where designers articulate requirements with prompts. Such a change could potentially revolutionize chip design by maximizing designer creativity and efficiency at scale.


Although general LLMs like GPTs can be adapted to a wide range of fields\cite{haozhaochatsim,adapt}, finetuning open-source LLMs is necessary to enhance the capability of democratized chip design process through domain customization, as it does in \emph{e.g.,} NVIDIA ChipNemo\cite{chipnemo} and academic community\cite{benchmarking, verigen, he2023chateda}. Compared to prior finetuned LLMs on Verilog code completion, however, as shown in Fig. \ref{fig:agent}, we believe a true \textcolor{blue}{Chip Design Agent} based on LLM should be able to act as an interface between human developers by \textcolor{blue}{conducting natural language to Verilog code translation, architecture design, feedback analysis from EDA tools and Verilog/script remodification}, besides Verilog completion as in prior code-gen LLMs. Such an LLM agent works like a human programmer by interacting with EDA tools feedback to remodify the Verilog code and script.
Such an ambitious attempt to reach chip-design LLMs requires very high-quality hardware-specific datasets, which are very limited in the open-source community as shown in Fig.2, and the situation poses a significant challenge to the goal of finetuning chip-design LLMs. In general, there are three major obstacles to address before the successful finetuning of a chip-design LLM. 
First of all, high-volume hardware datasets including Verilog and hardware description are in demand, while the open-source Verilog codes are very scarce regarding the million-scale requirement of domain-specific LLM finetuning. As we can see from Fig. \ref{fig:scalinglaw}, it takes at least $10^{8}$ entries of code and labels to unleash the intelligence of a domain-specific LLM.

Second, in addition to data volume, the quality of hardware datasets is also important in terms of diversity and consistency. For diversity, there are at least three types of datasets that are required such as natural language and HDL description, EDA feedback and scripts. As to consistency,  for Verilog codes, a chip-design LLM needs well-aligned pairs of natural language description and Verilog code, which respectively correspond to input and desired output of a LLM.
Besides Verilog codes, error feedback from EDA tools and the corresponding information on Verilog/script modification should also be provided and aligned, so that it can understand and manipulate EDA Tool through scripts such as the SiliconCompiler\cite{Siliconcomplier} python script. 

To address the data scarcity issue that limits the capability of a chip-design LLM, this paper proposes a full-stack hardware design-data augmentation framework to solve the above challenges. First of all, we propose an automated design-data augmentation framework to generate large-scale hardware design data without human intervention. For the Verilog generation task, we proposed a Verilog program analysis method to achieve alignment between Verilog and natural language, which maps the Verilog AST(Abstract Syntax Tree) directly to natural language. For Verilog repair task, we use the feedback from 
Verilog semantic checker (\emph{i.e.,} yosys) paired with the provided right and wrong Verilog files to generate Verilog repairing dataset
, which aligns the generated Verilog to the response of EDA Tool. For EDA Tool script generation, we observe that although existing LLM (OpenAI GPT) cannot generate accurate EDA Script, it can understand EDA script well. Therefore, we leverage existing LLMs to produce natural language descriptions according to the EDA script. The generated natural language description and EDA script are paired and linked to the same entry of the dataset.

We evaluate our data augmentation framework by finetuning Llama-2 7B and 13B and test it on third-party benchmarks. The results show that the chip design model finetuned with our design-data framework show a higher pass rate over the baseline models(GPT3.5, General Code generation, Thakur et al.\cite{benchmarking}) in Verilog generation, Verilog Repair and EDA Script Generation. The contributions are listed below:
\begin{itemize}
    \item We design and implement a design-data augmentation framework for training Chip Design LLMs that can generate Verilog, EDA script, and coordinate EDA-flow by receiving only natural language design description. The framework facilitates the automatic generation of high-volume and high-quality datasets including aligned Verilog/EDA-script information. To align natural language and Verilog, we use the program analysis method to generate natural language descriptions, which increases the pass rate from 25.7\% to 45.7\% in 13B model compared with the naive data generation method(\emph{i.e.,} only code completion). 
    \item We use the data generated by the proposed framework to finetune Llama 2-7B, 13B, which improves from 31.4\% to 45.7\% in function pass rate compared with Thakur et al.\cite{benchmarking}. It has a decent pass rate close to GPT-3.5 and outperforms it in some tasks.  
    \item The evaluation also shows that the proposed data augmentation framework maintains the data distribution and enhances the Verilog generation quality over the baseline dataset with the same size. In evaluation, we obtain 3671k data from this framework, which can promote the pass rate from 25.7\% to 45.7\% when applied to a 13B model compared with the general data generation method(only code completion).
\end{itemize}
\section{Background \& Motivation}
\paragraph{Chip-design Large Language Model. } Large language models (LLMs) have emerged as a promising technique for chip design. Previous research has investigated the utilization of LLMs for the generation of Hardware Description Language (HDL) code, such as Verilog, from natural language descriptions of hardware modules\cite{benchmarking,lu2023rtllm,liu2023Verilogeval}. Progress has been made on tasks like generating complete Verilog code \cite{verigen}, general register-transfer level (RTL) synthesis \cite{blocklove2023chip}, and enhancing open-source LLMs for EDA tool script writing \cite{he2023chateda}. LLM-based EDA has also been applied to domains like quantum computing \cite{quantumllm}, in-memory computing \cite{cimllm}, hardware verification \cite{fixing,llmassertion,orenesvera2023rtl}, and AI accelerator design \cite{fu2023gpt4aigchip}.
These efforts demonstrate the great potential of LLMs for automating hardware generation. However, most rely on proprietary commercial models. Developing open-source, portable LLM solutions is crucial for widespread adoption. Tab. \ref{tab:workclass} summarizes recent methods for finetuning LLMs on hardware generation tasks. For example, NVIDIA finetuned a model called ChipNeMo using proprietary data, precluding full reproducibility \cite{chipnemo}. We aim to enable high-quality finetuning of open-source LLMs without reliance on private data or specialized hardware. 


\begin{table}[htbp]
\caption{Comparison of hardware generation large language models. }
\label{tab:workclass}
\resizebox{\linewidth}{!}{%
\begin{tabular}{|l|l|l|l|l|l|}
\hline
\textbf{Works} & \textbf{Target Task}                                                                                          & \textbf{\begin{tabular}[c]{@{}l@{}}Pre-Trained\\ Model\end{tabular}} & \textbf{\begin{tabular}[c]{@{}l@{}}Target \\ Language\end{tabular}}                & \textbf{Data}                                         & \textbf{\begin{tabular}[c]{@{}l@{}}Auto\\ Aug.\end{tabular}} \\ \hline
ChipNeMo\cite{chipnemo}     & \begin{tabular}[c]{@{}l@{}}Verilog\\ Generation\end{tabular}                                                  & Llama 2                                                                    & Verilog                                                                            &  Private                                                    &                                                             $\times$ \\ \hline
Thakur et al.\cite{benchmarking}        & \begin{tabular}[c]{@{}l@{}}Verilog\\ Completion\end{tabular}                                                  & CodeGen                                                              & Verilog                                                                            & \begin{tabular}[c]{@{}l@{}}Github\\ etc.\end{tabular} &                                                              $\times$\\ \hline
ChatEDA\cite{he2023chateda}        & \begin{tabular}[c]{@{}l@{}}EDA Script \\ Generation\end{tabular}                                              & Llama 2                                                              & \begin{tabular}[c]{@{}l@{}}ChatEDA\\ (Python DSL)\end{tabular}                     & Custom                                                &                                                              $\times$\\ \hline
Ours           & \begin{tabular}[c]{@{}l@{}}Verilog\\ Generation, \\ Verilog\\ Repair, \\ EDA Script\\ Generation\end{tabular} & Llama 2                                                              & \begin{tabular}[c]{@{}l@{}}Verilog, \\ SiliconCompiler\\ (Python DSL)\end{tabular} & \begin{tabular}[c]{@{}l@{}}Github\\ etc.\end{tabular} &                                                              $\checkmark$\\ \hline
\end{tabular}
}
\end{table}


\paragraph{Hardware Data Scarcity \& Large Language Model is Data-limited under some circumstance.}  As deep learning models have transitioned to the large language model era, the primary bottleneck has shifted from model architecture to available data\cite{model2data}. OpenAI demonstrated a "scaling law" where increasing model size alone plateaus, and performance gains depend on more data as 
 shown in Fig. \ref{fig:scalinglaw}. Data augmentation has thus become a key technique for improving models' generalization through exposure to a richer, more diverse training distribution\cite{dataaugpt}.
In hardware domains like Verilog, limited labeled data presents unique challenges for applying large language models (LLMs). As Fig. \ref{fig:codestatisticfile} shows, Verilog code repositories contain orders of magnitude fewer files than general languages. While sufficient for basic code completion, the scarcity of paired natural language descriptions prevents direct training for translation tasks. Additionally, lacking syntax analysis from Verilog checkers precludes LLMs from learning automatic error correction abilities.
EDA script generation faces similar data scarcity issues. For some applications, creating extensive labeled training datasets is the most labor-intensive process in machine learning development\cite{he2023chateda}. To address these hardware domain data limitations, we propose a comprehensive data augmentation framework. It can overcome the following challenges in previous works.


\paragraph{Challenge 1: Natural language Verilog generation requires strict alignment between Verilog and Natural language.} During the LLM finetuning process, the generated Verilog should be aligned with the input natural language description. However, in previous LLM finetuning process, only Verilog file finetune without any natural language comment may  cause pass rate decrease as shown in Tab. \ref{tab:evalgeneration}. Therefore, we use a Verilog program analysis to help natural language and Verilog alignment, which can generate corresponding natural language from Verilog file. The core part is natural language generation using Antlr4, which is a parser. Our framework provides Verilog file and grammar file then we can do program analysis on the abstract syntax tree. The program analysis stage uses Verilog Abstract Syntax Tree to compile each syntax node to natural language, which can obtain alignment natural language only with Verilog file. 
\paragraph{Challenge 2: Verilog repair requires strict alignment between Verilog and EDA Tool Error feedback.} Previous Chip-design LLM\cite{benchmarking} can not leverage EDA Tool feedback to correct wrong Verilog. To solve this problem, we use rule-based approach to generate  error Verilog file from right Verilog file and obtain the Verilog checker's(Yosys) output. This method can generate Verilog repair with EDA tool alignment from Verilog file.
\paragraph{Challenge 3: EDA script generation requires high alignment with EDA script and Natural language.}
Previous EDA script generation only uses script as the input of LLM\cite{he2023chateda}. However, for the lack of EDA Tool example script and the natural language description, LLM is hard to setup connections between natural language and EDA script, which nearly output wrong file. To solve this problem, we observe that existing LLM(Pre-trained LLM, \emph{e.g.} OpenAI GPT) can understand EDA script but can not generate EDA script. Therefore, we give the existing LLM right EDA script and get the corresponding natural language, which obtains a high quality dataset and leverages the LLMs' understanding ability.  

\section{Automated Design-Data Augmentation Framework}
\begin{figure}[htbp]
    \centering
    \includegraphics[width=\linewidth]{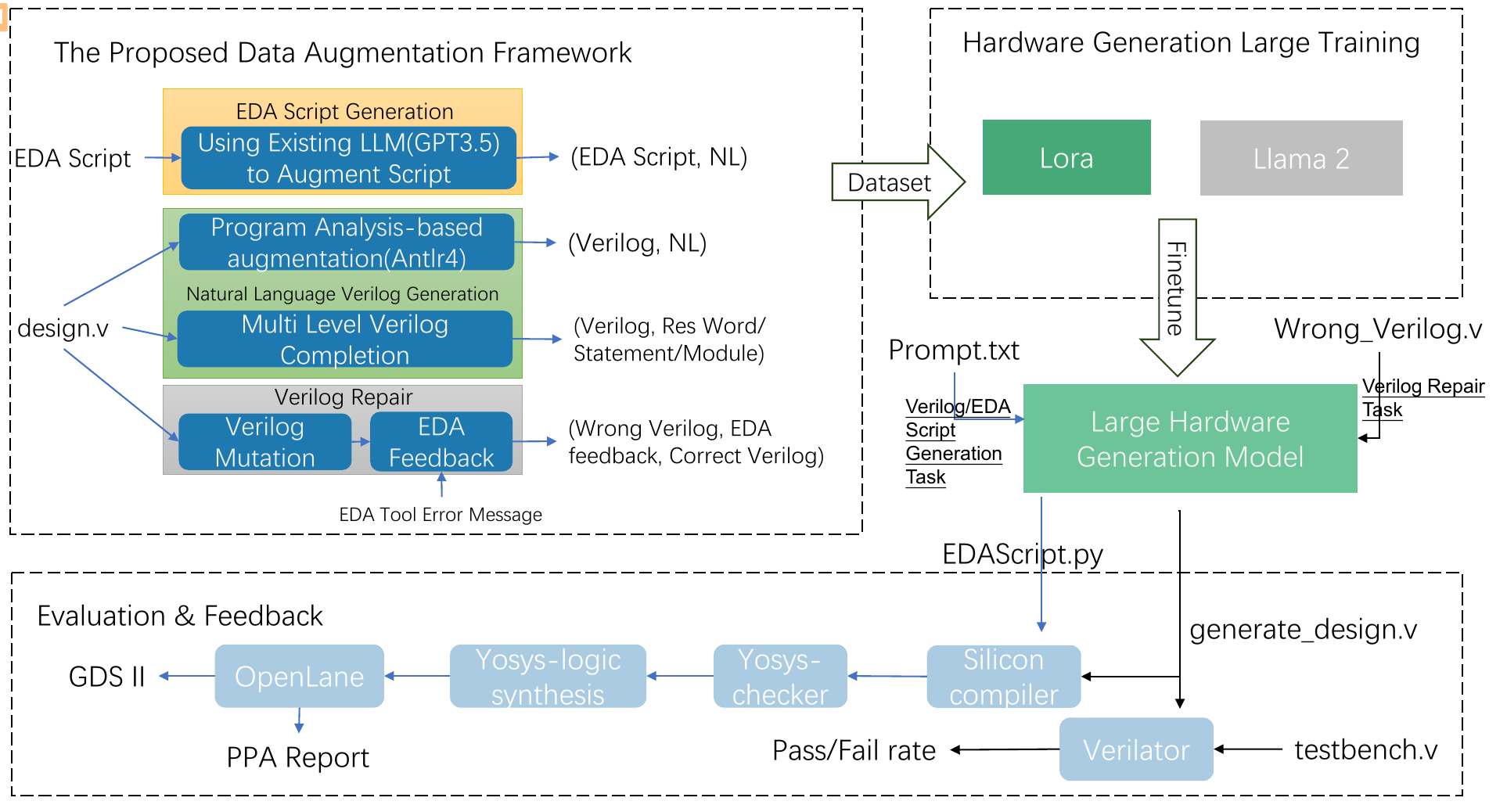}
    \caption{Overall workflow for hardware generation LLMs incorporating the proposed data augmentation framework. }
    \label{fig:overview}
\end{figure}

To develop an LLM capable of serving as an automated hardware generation EDA tool agent (Fig. \ref{fig:agent}), our proposed methodology involves the multi-stage data generation workflow depicted in Fig. \ref{fig:overview}. The training data generated through this process contains three key fields: 1) An instruction field that distinguishes the expected task (e.g.,code generation vs. error checking). 2) An input field containing the contextual information or prompt for the task. 3) An output field with the corresponding expected result.

\subsection{Data Augmentation for Verilog Generation}
To augment Verilog data for tasks beyond basic code completion, we propose a two-stage data augmentation process. It is well understood that LLMs tend to be strongly influenced by recent training examples\cite{dataorder}. Accordingly, our augmentation framework first exposes the model to larger quantities of less refined data to expand its initial knowledge base. This is followed by a second stage involving higher quality, more precisely targeted samples to refine the model's abilities.
\subsubsection{Basic Verilog Completion Augmentation Stage}
Code completion is a widely used data augmentation technique for sequence generation tasks. To strengthen an LLM's ability to predict Verilog code, we employ completion as the basic approach. Verilog completion can be formulated as using an initial token sequence $\{c_0, c_1,\cdots, c_{n-1}\}$  to predict the remaining tokens $\{c_{n},c_{n+1},\cdots, c_{m}\}$, where $c_i$ represents a character.
In our framework, completion examples are represented as: \{"instruct":"complete the next [level] of Verilog file.", "input":"[Existing Verilog]","output":"[Predict Verilog]"\}. The Verilog prediction data is separated into three coding granularity levels: module-level, sentence-level, and token-level. Module-level prediction uses the module header to generate the body. Sentence-level predicts the next statement given prior code ending in ';'. Token-level forecasts the subsequent token. A Verilog module with $i$ token, $j$ sentence can be divided into $1+j+i$ segments. 
As our evaluation in Tab.\ref{tab:evalgeneration} indicates, 
 code completion alone yields a small pass rate improvement from 22.9\% to 25.7\% (Tab. \ref{tab:evalgeneration}) compare with naive Llama 2, necessitating additional techniques to better align natural language and Verilog domains. Later framework stages introduce more sophisticated augmentation for broader capabilities.

\subsubsection{Natural language and Verilog Alignment - Using Program Analysis Rule}
To better align natural language with Verilog semantics, we develop a rule-based program analysis to translate Verilog code into structured natural language descriptions. The Verilog file is first parsed into an abstract syntax tree using ANTLR4. Translation rules are then applied to the syntax tree. For example, the rule module head translate \texttt{module x(input a, output reg b);} into "\texttt{The Verilog module with name [x] has one input [a] and one output [b]. The output is reg.}". This stage can be formulated as $Description=Rule(Verilog)$. Importantly, the ruleset does not capture full Verilog syntax. This mirrors how programmers typically use only core details when using  natural language to generate Verilog.
This stage generates training data in the format: D = \{ "instruct": "give me the Verilog module of this description.", "input":"[natural language]", "output":"[Verilog file]" \}.  Given a Verilog file with k translatable syntax structures, the dataset size increases linearly at $\mathcal{O}(k)$. This alignment stage can improve the LLM's pass rate from 25.7\% to 45.7\% (Tab. \ref{tab:evalgeneration}), which are similar with GPT-3.5 only with 13B weights. The rule-based approach effectively bridges the semantic gap between the natural language and the Verilog semantic.

\begin{figure}[htbp]
    \centering
    \includegraphics[width=\linewidth]{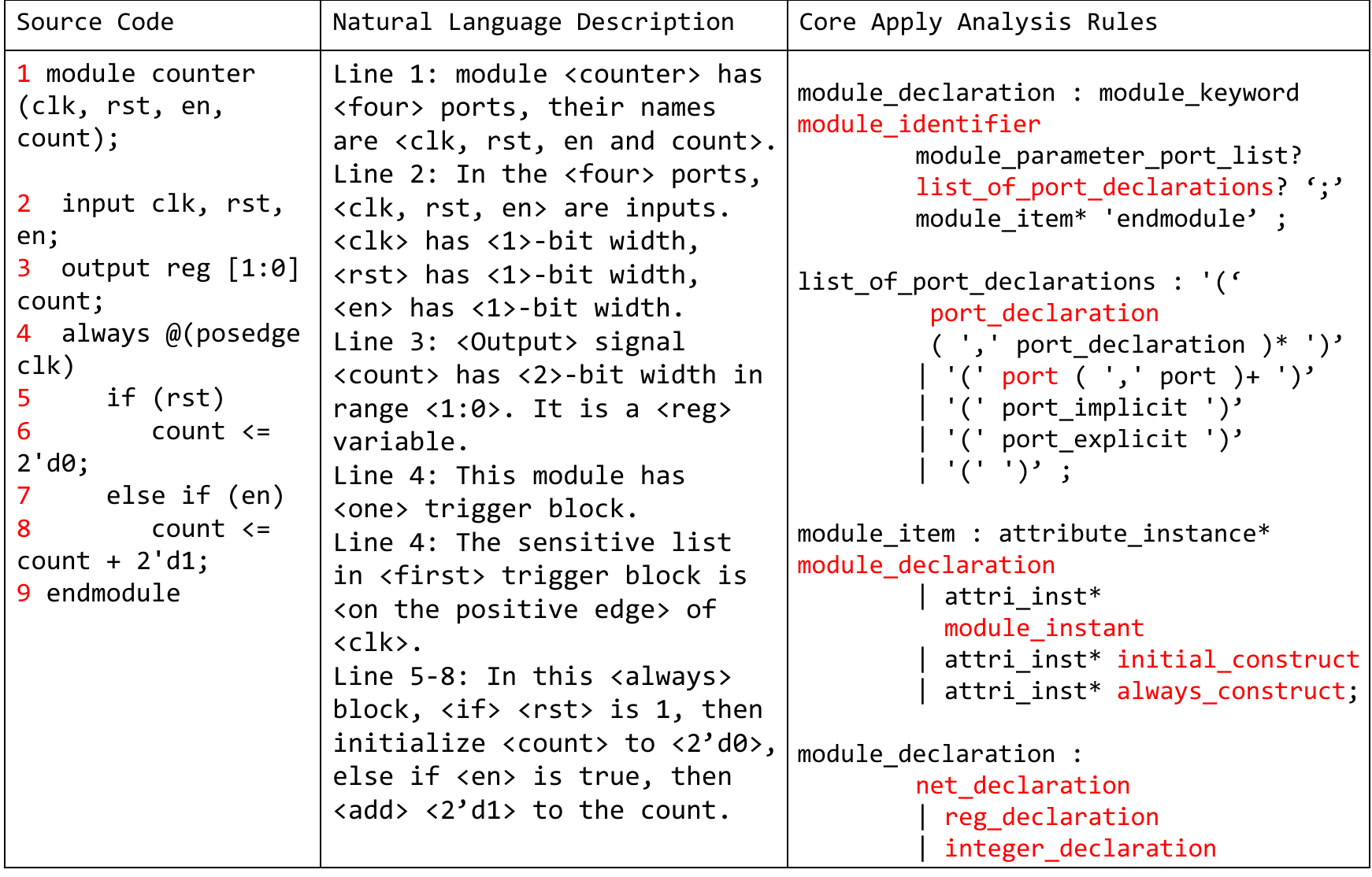}
    \caption{Natural Language Generation Using Program Analysis Rule. Core parts EBNF program analysis rules.}
    \label{fig:ebnf}
\end{figure}

To illustrate this technique in more detail, we conduct a case study using a sample Verilog module. As shown in Fig. \ref{fig:ebnf}, the module is first parsed into an abstract syntax tree using ANTLR. The example then apply rules to extract semantic information from the syntax structure. For example:

\begin{itemize}
    \item module \& port declaration: Compile modules' declarations into natural language. For example, "module counter(clk, rst, en, count)" is compiled to "module <counter> has <four> parts, their names are <clk, rst, en  and count>".
    \item always block declaration: Compile always block declaration into natural language. For example, in Fig. \ref{fig:ebnf}, the always block is tranlated into "The sensitive list in <first> trigger block is <on the positive edge> of <clk>.".
    \item variable declaration: Compile variable declaration into natural language. For example, in Fig. \ref{fig:ebnf}, data augmentation framework outputs "<Output> signal <count> has <2>-bit width in range <1:0>. It is a <reg> variable".
\end{itemize}

\subsection{Data Augmentation for Verilog Repair}
\subsubsection{Basic Verilog Code Repair Augmentation}\label{sec:vrepair}
When editing Verilog code, programmers may unintentionally introduce syntactic errors. A "Verilog repair" task aims to automatically correct such incorrect programs. However, training data for this task is challenging to obtain at scale as it requires incorporating realistic mistakes. We address this data scarcity issue through a rule-based targeted Verilog code masking approach. To construct samples mimicking errors, our framework programmatically masks tokens in correct Verilog programs. These tokens correspond to nodes in the Verilog syntax tree parsed by ANTLR4. In this task, the data augmentation framework generates  D = \{ "instruct": "give me correct Verilog according to the given wrong Verilog.", "input":"[wrong Verilog file]", "output":"[right Verilog file]" \}. For a Verilog file containing $x$ tokens, this approach yields $2x$ tokens worth of input-output pairs. 

To maintain high data quality, the number of changes made to any individual right Verilog module was limited to below five. A case is shown in Fig. \ref{fig:repair}, our algorithm changes the third column Verilog to the first column Verilog. The following lists key rules that were used to introduce targeted errors into correct Verilog code during the basic code repair data augmentation process:
\begin{itemize}[itemsep=2pt,topsep=0pt,parsep=0pt,partopsep=0pt,leftmargin=15pt]
    \item Generate Word Missing: Remove keywords, semicolon and operand.
    \item Generate Type Error: Change \texttt{wire} to \texttt{reg} or reverse on the abstract syntax tree.
    \item Generate Width Error: Add or sub the width value in \texttt{wire} and \texttt{reg} definitions.
    \item Generate Additional Word Error: Random add non-sense words in Verilog module.
    \item Generate Logic Error: Random remove logic condition statements in if statement.
\end{itemize}
\subsubsection{Verilog Code Repair Augmentation with EDA Tool Feedback. }  
\begin{figure}
    \centering
    \includegraphics[width=\linewidth]{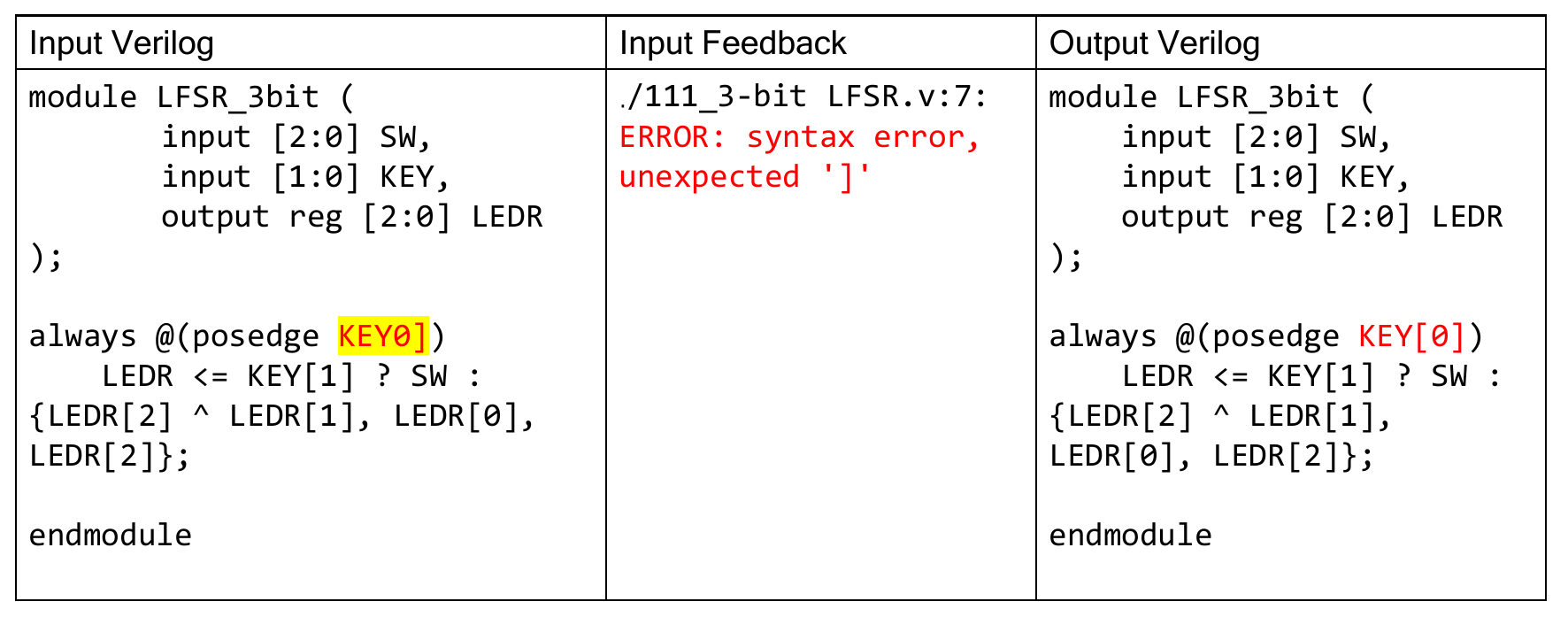}
    \caption{Our framework generated Verilog repair Data with EDA Tool.}
    \label{fig:repair}
\end{figure}

When generating Verilog code from natural language, syntactic or semantic errors can occur in the output. We observed that EDA tool feedback provides valuable information to improve code quality. During logic synthesis, Yosys (an ASIC logic synthesizer) checks Verilog syntax correctness and reports any errors. We leverage this capability by running syntax checks on the masked Verilog programs generated for "Verilog repair" task (Sec. \ref{sec:vrepair}). 
 Specifically, each masked program in Sec. \ref{sec:vrepair} is fed to Yosys to obtain error information. Our framework then pairs this feedback with the original incorrect code sample.  Concretely, as depicted in Fig. \ref{fig:repair}, the data takes the form:  D = \{ "instruct": "give me correct Verilog according to the given wrong Verilog.", "input":"[yosys
 info],[wrong Verilog file]", "output":"[right Verilog file]" \}. By incorporating EDA Tool diagnostics in this manner, our approach grounds Verilog generation in realistic tool constraints.

\subsection{Data Augmentation for EDA Tool Script Generation}\label{sec:edatool}

We selected the SiliconCompiler library as our baseline EDA tool, as it is a widely used open-source Python framework. However, the volume of existing sample scripts within SiliconCompiler's documentation is insufficient for largescale model finetuning. To address this, we propose using LLMs to augment the dataset while ensuring outputs adhere to the constraints of the SiliconCompiler domain.
Although direct LLM generation without domain knowledge risks producing scripts that are syntactically correct but semantically invalid, we observe that the existing LLM can understand SiliconCompiler's script and output a right natural language description. Therefore, we provided existing LLM(\emph{i.e.,} GPT-3.5) with around 200 examples of valid SiliconCompiler scripts in Python file format to obtain the dataset as shown in Equ. \ref{equ:edascript}. 
\begin{equation}\label{equ:edascript}
    GeneralLLM(\text{SiliconCompiler Script})=\text{Natural language Desc.}
\end{equation}
Surprisingly, with only these small samples(\emph{i.e.} 200), our approach outperformed baselines like GPT-3.5 and others one manitude which required far more training data.
For this stage, data augmentation framework generates training instances in the format:
D = \{ "instruct": "give me SiliconCompiler script.", "input":"[LLM generated description]", "output":"[SiliconCompiler script]" \}. This method relies on LLM tokens, where high computational costs limits its applicability for Verilog generation tasks.

\section{Implementation}
\paragraph{Dataset} This paragraph illustrates the dataset scale after implementing the proposed data augmentation method,  The augmented dataset scale is detailed in Tab \ref{tab:datasetscale}.
We trim the data that exceeds the maximum token length, as such data may introduce certain issues during model training. 
\begin{table}[htbp]
\caption{ Dataset Scale through Data Augmentation Framework. }
\label{tab:datasetscale}
\scalebox{0.8}{
\begin{tabular}{|l|l|l|}
\hline
\textbf{Task}                                                                        & \textbf{\begin{tabular}[c]{@{}l@{}}Output \\ Data Size\end{tabular}} & \textbf{\begin{tabular}[c]{@{}l@{}}Output \\ Data Number\end{tabular}} \\ \hline
\begin{tabular}[c]{@{}l@{}}Natural Language     \\ Verilog Generation\end{tabular} & 1784.24MB                                                            & ~124k                                                                  \\ \hline
Verilog Mask Completion                                                              & 2145.29MB                                                            & ~107k                                                                  \\ \hline
Verilog Debug                                                                        & 523.77MB                                                             & ~240k                                                                  \\ \hline
\begin{tabular}[c]{@{}l@{}}Verilog Word- \\ Level Completion\end{tabular}            & 21GB                                                                 & ~3700k                                                                 \\ \hline
\begin{tabular}[c]{@{}l@{}}Verilog Module-\\ Level Completion\end{tabular}           & 693MB                                                                    & 400k                                                                     \\ \hline
\begin{tabular}[c]{@{}l@{}}Verilog Statement-\\ Level Completion\end{tabular}        & 2.9GB                                                                  & 2388k                                                                   \\ \hline
\begin{tabular}[c]{@{}l@{}}Natural Language\\ EDA Script Generation\end{tabular}     & 301KB                                                              & 200                                                                   \\ \hline
\end{tabular}
}
\end{table}
\paragraph{Finetuning Large Language Model} We employ LoraNet\cite{hu2022lora} to finetune Llama2\cite{Llama2} on these datasets, based  on the Llama-recipes repository. The model comprises 40 hidden layers, maximum position embeddings of 2048, and is trained over a period of approximately 7 days.  Additionally, our methodology involves the following steps:
Step 1: Data collection from GitHub and HuggingFace datasets.
Step 2: Application of our proposed data augmentation framework.
Step 3: Fine-tuning of the pre-trained LLama 2 large language model.
Step 4: Utilization of the  benchmark for evaluating data augmentation performance.
Our evaluation results demonstrate the effectiveness of the proposed data augmentation method.
\subsection{Evaluation Setup}

\paragraph{Finetune Environment} We harness the extensive language library \texttt{transformers} and the distributed framework \texttt{deepspeed}, both constructed on the foundational framework PyTorch. Our infrastructure includes a cluster with 8 NVIDIA A100 GPUs, each equipped with 40 GB of memory, and an Intel Xeon CPU. 

\paragraph{Baseline}  We choose Thakur et al.\cite{benchmarking,verigen} (SOTA Open Source Verilog Generation Model), GPT3.5 as the primary baseline model to access the capabilities of our model. Here are the utilized models for comparsion. ChatEDA\cite{he2023chateda} is not open-source yet, hence it is excluded from the comparison.
\begin{itemize}
    \item \textbf{Llama 2-PT 13B} LLama 2 is an open-source large language model developed by Meta.
    \item \textbf{Llama 2-FT (Ours) 7B} We employ the proposed data augmentation workflow to fine-tune LLama 2-7B.
    \item \textbf{Llama 2-FT (Ours) 13B} We employ the proposed data augmentation workflow to fine-tune LLama 2-13B.
    \item \textbf{Llama 2-FT (General Aug) 13B} We fine-tune Llama 2-13B  only with code completion as an ablation study baseline. 
    \item \textbf{GPT 3.5} The most widely used LLM from OpenAI. The weight parameters are not publicly available.
    \item \textbf{Thakur et al.} A Verilog generation large language model based on Codegen-16B.
\end{itemize}
\paragraph{Benchmark \& EDA Environment} For the LLM inference, we set temperature to 0.1(less creative) and beams to 4(larger search space). To ensure a fair comparison, we select a subset of Natural language benchmark Thakur et al.\cite{benchmarking} and RTLLM\cite{lu2023rtllm} as our benchmark. We employ VCS as the functional simulator. The backend of SiliconCompiler operates on openlane, utilizing the SkyWater 130nm Process Design Kit (PDK). 

\subsection{Evaluation Result}
\begin{figure}
    \centering
    \includegraphics[width=\linewidth]{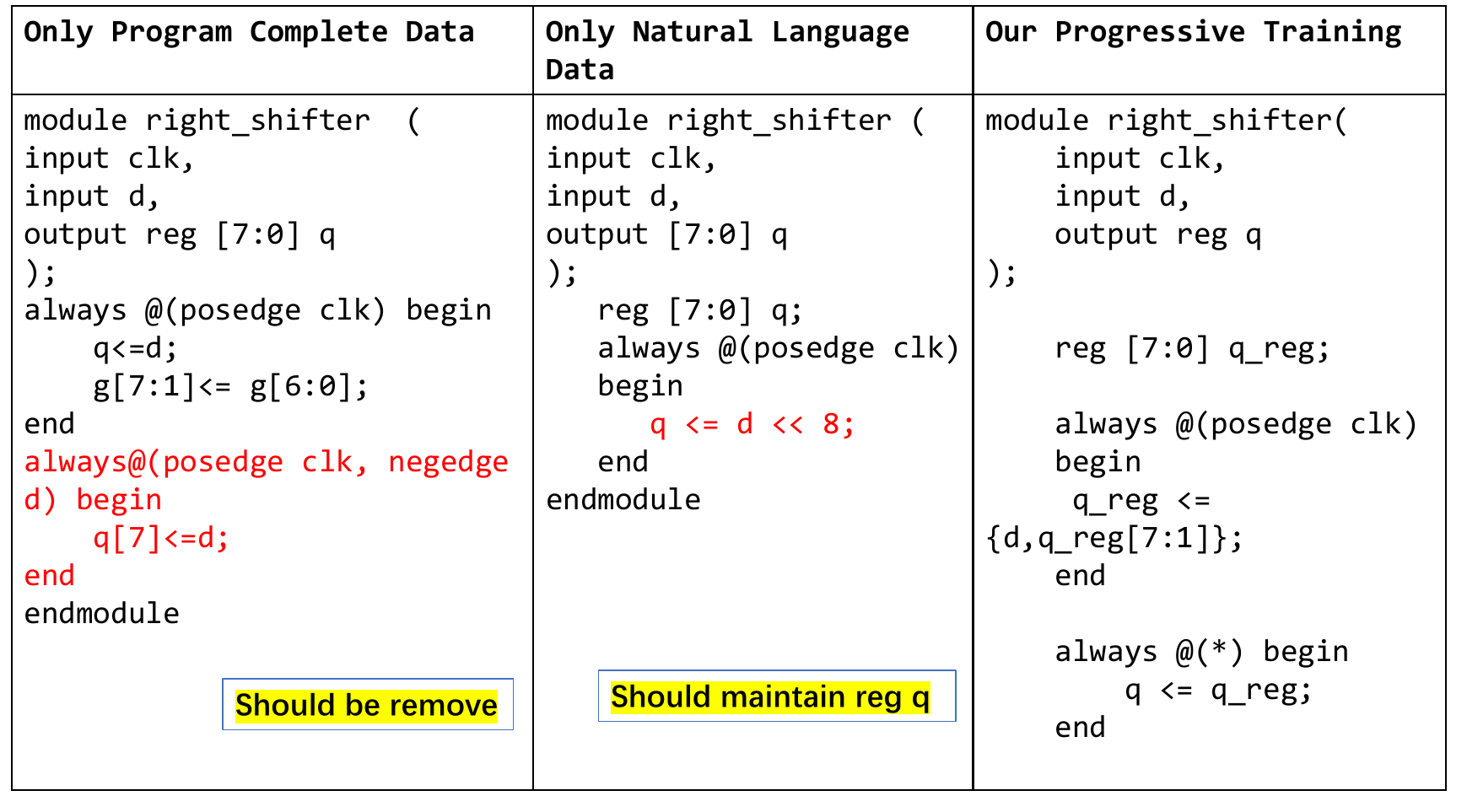}
    \caption{Ablation Study for the Data Augmentation Framework.}
    \label{fig:ablationstudy}
\end{figure}

\begin{table}[htbp]
\caption{Evaluation for Verilog repair using RTLLM\cite{lu2023rtllm} benchmark. Syntax  represents the number of Verilog code generated by LLM with syntax errors under pass@5. Function represents   testbench pass rate of the best-performing Verilog code under pass@5.}
\label{tab:evalrepair}
\resizebox{\linewidth}{!}{%
\begin{tabular}{|c|cc|cc|cc|cc|}
\hline
\multirow{2}{*}{\textbf{Benchmark}} &
  \multicolumn{2}{c|}{ours-13B} &
  \multicolumn{2}{c|}{ours-7B} &
  \multicolumn{2}{c|}{GPT3.5} &
  \multicolumn{2}{c|}{Llama2-13B} \\ \cline{2-9} 
 &
  \multicolumn{1}{c|}{syntax} &
  function &
  \multicolumn{1}{c|}{syntax} &
  function &
  \multicolumn{1}{c|}{syntax} &
  function &
  \multicolumn{1}{c|}{syntax} &
  function \\ \hline
\textbf{accu} &
  \multicolumn{1}{c|}{0} &
  100\% &
  \multicolumn{1}{c|}{0} &
  100\% &
  \multicolumn{1}{c|}{0} &
  100\% &
  \multicolumn{1}{c|}{4} &
  0\% \\ \hline
\textbf{adder\_8bit} &
  \multicolumn{1}{c|}{0} &
  0\% &
  \multicolumn{1}{c|}{0} &
  0\% &
  \multicolumn{1}{c|}{1} &
  100\% &
  \multicolumn{1}{c|}{5} &
  0\% \\ \hline
\textbf{adder\_16bit} &
  \multicolumn{1}{c|}{0} &
  0\% &
  \multicolumn{1}{c|}{0} &
  0\% &
  \multicolumn{1}{c|}{2} &
  100\% &
  \multicolumn{1}{c|}{5} &
  0\% \\ \hline
\textbf{adder\_32bit} &
  \multicolumn{1}{c|}{0} &
  0\% &
  \multicolumn{1}{c|}{0} &
  0\% &
  \multicolumn{1}{c|}{4} &
  0\% &
  \multicolumn{1}{c|}{5} &
  0\% \\ \hline
\textbf{adder\_64bit} &
  \multicolumn{1}{c|}{0} &
  0\% &
  \multicolumn{1}{c|}{0} &
  0\% &
  \multicolumn{1}{c|}{5} &
  0\% &
  \multicolumn{1}{c|}{5} &
  0\% \\ \hline
\textbf{multi\_16bit} &
  \multicolumn{1}{c|}{0} &
  100\% &
  \multicolumn{1}{c|}{0} &
  100\% &
  \multicolumn{1}{c|}{5} &
  0\% &
  \multicolumn{1}{c|}{3} &
  0\% \\ \hline
\textbf{multi\_pipe\_4bit} &
  \multicolumn{1}{c|}{0} &
  0\% &
  \multicolumn{1}{c|}{0} &
  0\% &
  \multicolumn{1}{c|}{5} &
  0\% &
  \multicolumn{1}{c|}{1} &
  100\% \\ \hline
\textbf{multi\_pipe\_8bit} &
  \multicolumn{1}{c|}{0} &
  100\% &
  \multicolumn{1}{c|}{3} &
  0\% &
  \multicolumn{1}{c|}{5} &
  0\% &
  \multicolumn{1}{c|}{1} &
  0\% \\ \hline
\textbf{multi\_booth} &
  \multicolumn{1}{c|}{0} &
  100\% &
  \multicolumn{1}{c|}{0} &
  0\% &
  \multicolumn{1}{c|}{4} &
  0\% &
  \multicolumn{1}{c|}{0} &
  0\% \\ \hline
\textbf{div\_16bit} &
  \multicolumn{1}{c|}{0} &
  100\% &
  \multicolumn{1}{c|}{5} &
  0\% &
  \multicolumn{1}{c|}{4} &
  0\% &
  \multicolumn{1}{c|}{0} &
  100\% \\ \hline
\textbf{radix2\_div} &
  \multicolumn{1}{c|}{0} &
  100\% &
  \multicolumn{1}{c|}{0} &
  0\% &
  \multicolumn{1}{c|}{5} &
  0\% &
  \multicolumn{1}{c|}{1} &
  0\% \\ \hline
\textbf{Johnson\_Counter} &
  \multicolumn{1}{c|}{0} &
  100\% &
  \multicolumn{1}{c|}{0} &
  100\% &
  \multicolumn{1}{c|}{0} &
  97\% &
  \multicolumn{1}{c|}{3} &
  100\% \\ \hline
\textbf{right\_shifter} &
  \multicolumn{1}{c|}{0} &
  100\% &
  \multicolumn{1}{c|}{0} &
  100\% &
  \multicolumn{1}{c|}{0} &
  0\% &
  \multicolumn{1}{c|}{0} &
  0\% \\ \hline
\textbf{mux} &
  \multicolumn{1}{c|}{0} &
  100\% &
  \multicolumn{1}{c|}{0} &
  100\% &
  \multicolumn{1}{c|}{0} &
  100\% &
  \multicolumn{1}{c|}{5} &
  0\% \\ \hline
\textbf{counter\_12} &
  \multicolumn{1}{c|}{0} &
  100\% &
  \multicolumn{1}{c|}{0} &
  100\% &
  \multicolumn{1}{c|}{4} &
  0\% &
  \multicolumn{1}{c|}{5} &
  0\% \\ \hline
\textbf{freq\_div} &
  \multicolumn{1}{c|}{0} &
  0\% &
  \multicolumn{1}{c|}{0} &
  0\% &
  \multicolumn{1}{c|}{3} &
  0\% &
  \multicolumn{1}{c|}{0} &
  0\% \\ \hline
\textbf{signal\_generator} &
  \multicolumn{1}{c|}{0} &
  100\% &
  \multicolumn{1}{c|}{2} &
  0\% &
  \multicolumn{1}{c|}{5} &
  0\% &
  \multicolumn{1}{c|}{5} &
  0\% \\ \hline
\textbf{serial2parallel} &
  \multicolumn{1}{c|}{0} &
  100\% &
  \multicolumn{1}{c|}{0} &
  100\% &
  \multicolumn{1}{c|}{3} &
  100\% &
  \multicolumn{1}{c|}{2} &
  0\% \\ \hline
\textbf{parallel2serial} &
  \multicolumn{1}{c|}{0} &
  0\% &
  \multicolumn{1}{c|}{0} &
  0\% &
  \multicolumn{1}{c|}{5} &
  0\% &
  \multicolumn{1}{c|}{5} &
  0\% \\ \hline
\textbf{pulse\_detect} &
  \multicolumn{1}{c|}{0} &
  0\% &
  \multicolumn{1}{c|}{0} &
  0\% &
  \multicolumn{1}{c|}{2} &
  0\% &
  \multicolumn{1}{c|}{4} &
  0\% \\ \hline
\textbf{edge\_detect} &
  \multicolumn{1}{c|}{0} &
  100\% &
  \multicolumn{1}{c|}{0} &
  100\% &
  \multicolumn{1}{c|}{2} &
  100\% &
  \multicolumn{1}{c|}{5} &
  0\% \\ \hline
\textbf{fsm} &
  \multicolumn{1}{c|}{0} &
  100\% &
  \multicolumn{1}{c|}{0} &
  100\% &
  \multicolumn{1}{c|}{4} &
  0\% &
  \multicolumn{1}{c|}{0} &
  0\% \\ \hline
\textbf{width\_8to16} &
  \multicolumn{1}{c|}{0} &
  100\% &
  \multicolumn{1}{c|}{0} &
  100\% &
  \multicolumn{1}{c|}{3} &
  30\% &
  \multicolumn{1}{c|}{4} &
  0\% \\ \hline
\textbf{traffic\_light} &
  \multicolumn{1}{c|}{0} &
  100\% &
  \multicolumn{1}{c|}{0} &
  100\% &
  \multicolumn{1}{c|}{5} &
  0\% &
  \multicolumn{1}{c|}{0} &
  0\% \\ \hline
\textbf{calendar} &
  \multicolumn{1}{c|}{0} &
  100\% &
  \multicolumn{1}{c|}{0} &
  100\% &
  \multicolumn{1}{c|}{2} &
  100\% &
  \multicolumn{1}{c|}{2} &
  0\% \\ \hline
\textbf{RAM} &
  \multicolumn{1}{c|}{0} &
  100\% &
  \multicolumn{1}{c|}{0} &
  100\% &
  \multicolumn{1}{c|}{1} &
  100\% &
  \multicolumn{1}{c|}{3} &
  0\% \\ \hline
\textbf{asyn\_fifo} &
  \multicolumn{1}{c|}{0} &
  100\% &
  \multicolumn{1}{c|}{5} &
  0\% &
  \multicolumn{1}{c|}{5} &
  0\% &
  \multicolumn{1}{c|}{2} &
  0\% \\ \hline
\textbf{alu} &
  \multicolumn{1}{c|}{0} &
  100\% &
  \multicolumn{1}{c|}{0} &
  100\% &
  \multicolumn{1}{c|}{5} &
  0\% &
  \multicolumn{1}{c|}{5} &
  0\% \\ \hline
\textbf{pe} &
  \multicolumn{1}{c|}{0} &
  100\% &
  \multicolumn{1}{c|}{0} &
  100\% &
  \multicolumn{1}{c|}{1} &
  100\% &
  \multicolumn{1}{c|}{5} &
  0\% \\ \hline
\textbf{success rate} &
  \multicolumn{1}{l|}{} &
  \multicolumn{1}{l|}{72.40\%} &
  \multicolumn{1}{l|}{} &
  \multicolumn{1}{l|}{51.70\%} &
  \multicolumn{1}{l|}{} &
  \multicolumn{1}{l|}{34.50\%} &
  \multicolumn{1}{l|}{} &
  \multicolumn{1}{l|}{10.30\%} \\ \hline
\end{tabular}%
}
\end{table}

\begin{table}[htbp]
\caption{Evaluation for SiliconCompiler script generation. syn represents the iterations needed to generate SiliconCompiler scripts with correct syntax, while func represents the iterations needed to generate SiliconCompiler scripts with correct function under pass@10.}
\label{tab:evalscript}
\resizebox{\linewidth}{!}{%
\begin{tabular}{|c|cc|cc|cc|cc|cc|}
\hline
\multirow{2}{*}{\textbf{benchmark}} &
  \multicolumn{2}{c|}{GPT3.5} &
  \multicolumn{2}{c|}{Thakur et al.\cite{benchmarking}} &
  \multicolumn{2}{c|}{Ours-7B} &
  \multicolumn{2}{c|}{LLama2-13B} &
  \multicolumn{2}{c|}{Ours-13B} \\ \cline{2-11} 
 &
  \multicolumn{1}{c|}{syn.} &
  func. &
  \multicolumn{1}{c|}{syn.} &
  func. &
  \multicolumn{1}{c|}{syn.} &
  func. &
  \multicolumn{1}{c|}{syn.} &
  func. &
  \multicolumn{1}{c|}{syn.} &
  func. \\ \hline
Basic &
  \multicolumn{1}{c|}{8} &
  9 &
  \multicolumn{1}{c|}{\textgreater{}10} &
  \textgreater{}10 &
  \multicolumn{1}{c|}{1} &
  1 &
  \multicolumn{1}{c|}{\textgreater{}10} &
  \textgreater{}10 &
  \multicolumn{1}{c|}{1} &
  1 \\ \hline
Layout &
  \multicolumn{1}{c|}{9} &
  10 &
  \multicolumn{1}{c|}{\textgreater{}10} &
  \textgreater{}10 &
  \multicolumn{1}{c|}{1} &
  1 &
  \multicolumn{1}{c|}{\textgreater{}10} &
  \textgreater{}10 &
  \multicolumn{1}{c|}{1} &
  1 \\ \hline
Clock Period &
  \multicolumn{1}{c|}{10} &
  \textgreater{}10 &
  \multicolumn{1}{c|}{\textgreater{}10} &
  \textgreater{}10 &
  \multicolumn{1}{c|}{1} &
  1 &
  \multicolumn{1}{c|}{\textgreater{}10} &
  \textgreater{}10 &
  \multicolumn{1}{c|}{1} &
  1 \\ \hline
Core Area &
  \multicolumn{1}{c|}{\textgreater{}10} &
  \textgreater{}10 &
  \multicolumn{1}{c|}{\textgreater{}10} &
  \textgreater{}10 &
  \multicolumn{1}{c|}{1} &
  1 &
  \multicolumn{1}{c|}{\textgreater{}10} &
  \textgreater{}10 &
  \multicolumn{1}{c|}{1} &
  1 \\ \hline
Mixed &
  \multicolumn{1}{c|}{\textgreater{}10} &
  \textgreater{}10 &
  \multicolumn{1}{c|}{\textgreater{}10} &
  \textgreater{}10 &
  \multicolumn{1}{c|}{2} &
  2 &
  \multicolumn{1}{c|}{\textgreater{}10} &
  \textgreater{}10 &
  \multicolumn{1}{c|}{2} &
  2 \\ \hline
\textbf{avg pass@k} &
  \multicolumn{1}{c|}{\textgreater{}8} &
  \textgreater{}9 &
  \multicolumn{1}{c|}{\textgreater{}10} &
  \textgreater{}10 &
  \multicolumn{1}{c|}{1} &
  1 &
  \multicolumn{1}{c|}{\textgreater{}10} &
  \textgreater{}10 &
  \multicolumn{1}{c|}{1} &
  1 \\ \hline
\end{tabular}%
}
\end{table}
\begin{table*}[]
\caption{Evaluation for Verilog Generation. Every cell in Thakur et al. benchmark\cite{benchmarking} consists the result of three  prompt levels(low/middle/high). Syntax  represents the number of Verilog code generated by LLM with syntax errors under pass@5. Function represents   testbench pass rate of the best-performing Verilog code under pass@5.}
\label{tab:evalgeneration}
\resizebox{\textwidth}{!}{%
\begin{tabular}{|cc|cc|cc|cc|cc|cc|cc|}
\hline
\multicolumn{2}{|c|}{\textbf{benchmark}} &
  \multicolumn{2}{c|}{GPT3.5} &
  \multicolumn{2}{c|}{Ours-7B} &
  \multicolumn{2}{c|}{Ours-13B} &
  \multicolumn{2}{c|}{Thakur et al.\cite{benchmarking}} &
  \multicolumn{2}{c|}{Llama2-13B} &
  \multicolumn{2}{c|}{LLama2-General Aug.} \\ \hline
\multicolumn{1}{|c|}{\multirow{18}{*}{Thakur et al.}} &
  Name &
  \multicolumn{1}{c|}{syntax} &
  function &
  \multicolumn{1}{c|}{syntax} &
  function &
  \multicolumn{1}{c|}{syntax} &
  function &
  \multicolumn{1}{c|}{syntax} &
  function &
  \multicolumn{1}{c|}{syntax} &
  function &
  \multicolumn{1}{c|}{syntax} &
  function \\ \cline{2-14} 
\multicolumn{1}{|c|}{} &
  basic1 &
  \multicolumn{1}{c|}{0/0/0} &
  100\%/100\%/100\% &
  \multicolumn{1}{c|}{0/0/0} &
  100\%/100\%/100\% &
  \multicolumn{1}{c|}{0/0/0} &
  100\%/100\%/100\% &
  \multicolumn{1}{c|}{0/0/0} &
  100\%/100\%/100\% &
  \multicolumn{1}{c|}{0/0/0} &
  100\%/100\%/100\% &
  \multicolumn{1}{c|}{0/0/0} &
  100\%/100\%/100\% \\ \cline{2-14} 
\multicolumn{1}{|c|}{} &
  basic2 &
  \multicolumn{1}{c|}{0/0/0} &
  100\%/100\%/100\% &
  \multicolumn{1}{c|}{0/0/0} &
  100\%/100\%/100\% &
  \multicolumn{1}{c|}{0/0/0} &
  100\%/100\%/100\% &
  \multicolumn{1}{c|}{0/0/0} &
  100\%/100\%/100\% &
  \multicolumn{1}{c|}{0/0/0} &
  100\%/100\%/100\% &
  \multicolumn{1}{c|}{0/5/0} &
  100\%/0\%/100\% \\ \cline{2-14} 
\multicolumn{1}{|c|}{} &
  basic3 &
  \multicolumn{1}{c|}{0/0/0} &
  12.5\%/37.5\%/100\% &
  \multicolumn{1}{c|}{0/0/0} &
  50\%/50\%/50\% &
  \multicolumn{1}{c|}{0/0/0} &
  25\%/50\%/50\% &
  \multicolumn{1}{c|}{0/0/0} &
  12.5\%/0\%/100\% &
  \multicolumn{1}{c|}{0/0/0} &
  0\%/12.5\%/100\% &
  \multicolumn{1}{c|}{0/0/0} &
  12.5\%/12.5\%/100\% \\ \cline{2-14} 
\multicolumn{1}{|c|}{} &
  basic4 &
  \multicolumn{1}{c|}{0/1/1} &
  100\%/100\%/100\% &
  \multicolumn{1}{c|}{0/0/0} &
  100\%/100\%/100\% &
  \multicolumn{1}{c|}{0/0/0} &
  100\%/100\%100\% &
  \multicolumn{1}{c|}{0/0/0} &
  0\%/100\%/100\% &
  \multicolumn{1}{c|}{0/0/0} &
  100\%/100\%/100\% &
  \multicolumn{1}{c|}{3/1/0} &
  100\%/100\%/100\% \\ \cline{2-14} 
\multicolumn{1}{|c|}{} &
  intermediate1 &
  \multicolumn{1}{c|}{0/0/0} &
  100\%/100\%/100\% &
  \multicolumn{1}{c|}{0/0/0} &
  25\%/25\%/25\% &
  \multicolumn{1}{c|}{2/0/0} &
  25\%/100\%/100\% &
  \multicolumn{1}{c|}{0/0/0} &
  100\%/100\%/0\% &
  \multicolumn{1}{c|}{0/5/5} &
  100\%/0\%/0\% &
  \multicolumn{1}{c|}{0/0/0} &
  75\%/75\%/100\% \\ \cline{2-14} 
\multicolumn{1}{|c|}{} &
  intermediate2 &
  \multicolumn{1}{c|}{0/0/0} &
  100\%/100\%/100\% &
  \multicolumn{1}{c|}{0/0/0} &
  68.4\%/68.4\%/100\% &
  \multicolumn{1}{c|}{0/0/0} &
  100\%/73.6\%/100\% &
  \multicolumn{1}{c|}{0/0/0} &
  0\%/0\%100\% &
  \multicolumn{1}{c|}{0/0/0} &
  0\%/0\%/100\% &
  \multicolumn{1}{c|}{0/0/0} &
  0\%/0\%/100\% \\ \cline{2-14} 
\multicolumn{1}{|c|}{} &
  intermediate3 &
  \multicolumn{1}{c|}{5/5/4} &
  0\%/0\%/6.25\% &
  \multicolumn{1}{c|}{5/5/5} &
  0\%/0\%/0\% &
  \multicolumn{1}{c|}{1/0/0} &
  11.7\%/11.7\%/11.7\% &
  \multicolumn{1}{c|}{0/0/0} &
  0\%/0\%/6.25\% &
  \multicolumn{1}{c|}{0/0/0} &
  0\%/0\%/0\% &
  \multicolumn{1}{c|}{5/5/5} &
  0\%/0\%/0\% \\ \cline{2-14} 
\multicolumn{1}{|c|}{} &
  intermediate4 &
  \multicolumn{1}{c|}{0/0/0} &
  0\%/0\%/0\% &
  \multicolumn{1}{c|}{0/0/0} &
  50\%/62.5\%/100\% &
  \multicolumn{1}{c|}{0/0/0} &
  37.5\%/62.5\%/75\% &
  \multicolumn{1}{c|}{1/0/0} &
  0\%/0\%/0\% &
  \multicolumn{1}{c|}{5/2/2} &
  0\%/0\%/0\% &
  \multicolumn{1}{c|}{0/0/0} &
  0\%/12.5\%/12.5\% \\ \cline{2-14} 
\multicolumn{1}{|c|}{} &
  intermediate5 &
  \multicolumn{1}{c|}{0/0/0} &
  100\%/100\%/100\% &
  \multicolumn{1}{c|}{0/0/0} &
  30\%/30\%/30\% &
  \multicolumn{1}{c|}{0/0/0} &
  10\%/60\%/60\% &
  \multicolumn{1}{c|}{0/0/0} &
  0\%/0\%/0\% &
  \multicolumn{1}{c|}{3/0/0} &
  0\%/20\%/60\% &
  \multicolumn{1}{c|}{0/5/5} &
  10\%/0\%/0\% \\ \cline{2-14} 
\multicolumn{1}{|c|}{} &
  intermediate6 &
  \multicolumn{1}{c|}{0/0/0} &
  0\%/0\%/0\% &
  \multicolumn{1}{c|}{5/5/0} &
  0\%/0\%/100\% &
  \multicolumn{1}{c|}{5/5/0} &
  0\%/0\%/100\% &
  \multicolumn{1}{c|}{0/0/0} &
  0\%/0\%/0\% &
  \multicolumn{1}{c|}{5/5/5} &
  0\%/0\%/0\% &
  \multicolumn{1}{c|}{5/5/5} &
  0\%/0\%/0\% \\ \cline{2-14} 
\multicolumn{1}{|c|}{} &
  intermediate7 &
  \multicolumn{1}{c|}{1/1/1} &
  100\%/100\%/100\% &
  \multicolumn{1}{c|}{0/0/0} &
  100\%/100\%/100\% &
  \multicolumn{1}{c|}{3/5/0} &
  100\%/100\%/100\% &
  \multicolumn{1}{c|}{1/0/0} &
  6\%/6\%/100\% &
  \multicolumn{1}{c|}{2/2/0} &
  0\%/0\%/0\% &
  \multicolumn{1}{c|}{0/1/0} &
  6\%/6\%/100\% \\ \cline{2-14} 
\multicolumn{1}{|c|}{} &
  intermediate8 &
  \multicolumn{1}{c|}{0/0/0} &
  100\%/100\%/100\% &
  \multicolumn{1}{c|}{0/0/5} &
  37.5\%/37.5\%/0\% &
  \multicolumn{1}{c|}{0/0/0} &
  25\%/80\%/80\% &
  \multicolumn{1}{c|}{2/4/5} &
  0\%/0\%/0\% &
  \multicolumn{1}{c|}{0/0/0} &
  62.5\%/62.5\%/62.5\% &
  \multicolumn{1}{c|}{0/0/5} &
  0\%/0\%/0\% \\ \cline{2-14} 
\multicolumn{1}{|c|}{} &
  advanced1 &
  \multicolumn{1}{c|}{3/1/1} &
  100\%/100\%/100\% &
  \multicolumn{1}{c|}{0/0/0} &
  49.9\%/51.5\%/75\% &
  \multicolumn{1}{c|}{0/0/0} &
  100\%/49.9\%/51.5\% &
  \multicolumn{1}{c|}{1/2/5} &
  25\%/25\%/0\% &
  \multicolumn{1}{c|}{1/1/0} &
  25.2\%/25.2\%/0.19\% &
  \multicolumn{1}{c|}{1/5/0} &
  50\%/0\%/25\% \\ \cline{2-14} 
\multicolumn{1}{|c|}{} &
  advanced2 &
  \multicolumn{1}{c|}{0/0/0} &
  92.8\%/92.8\%/92.8\% &
  \multicolumn{1}{c|}{0/0/0} &
  92.8\%/100\%/100\% &
  \multicolumn{1}{c|}{0/0/5} &
  92.8\%/100\%/100\% &
  \multicolumn{1}{c|}{0/0/0} &
  92.8\%/100\%/92.8\% &
  \multicolumn{1}{c|}{0/5/5} &
  0\%/0\%/0\% &
  \multicolumn{1}{c|}{0/0/0} &
  7\%/14\%/93\% \\ \cline{2-14} 
\multicolumn{1}{|c|}{} &
  advanced3 &
  \multicolumn{1}{c|}{0/0/1} &
  83\%/83\%/83\% &
  \multicolumn{1}{c|}{0/0/0} &
  100\%/100\%/100\% &
  \multicolumn{1}{c|}{0/0/5} &
  100\%/100\%/100\% &
  \multicolumn{1}{c|}{0/0/0} &
  100\%/0\%/100\% &
  \multicolumn{1}{c|}{1/0/0} &
  0\%/0\%/100\% &
  \multicolumn{1}{c|}{0/0/0} &
  66\%/66\%/100\% \\ \cline{2-14} 
\multicolumn{1}{|c|}{} &
  advanced4 &
  \multicolumn{1}{c|}{0/0/0} &
  62.5\%/62.5\%/62.5\% &
  \multicolumn{1}{c|}{0/0/0} &
  100\%/100\%/100\% &
  \multicolumn{1}{c|}{5/5/5} &
  100\%/100\%/100\% &
  \multicolumn{1}{c|}{1/0/0} &
  100\%/100\%/100\% &
  \multicolumn{1}{c|}{0/0/0} &
  12.5\%/12.5\%/50\% &
  \multicolumn{1}{c|}{0/0/0} &
  37.5\%/50\%/50\% \\ \cline{2-14} 
\multicolumn{1}{|c|}{} &
  advanced5 &
  \multicolumn{1}{c|}{0/0/0} &
  37.5\%/37.5\%/100\% &
  \multicolumn{1}{c|}{5/0/0} &
  0\%/50\%/100\% &
  \multicolumn{1}{c|}{5/0/0} &
  37.5\%/75\%/100\% &
  \multicolumn{1}{c|}{0/0/5} &
  0\%/75\%/0\% &
  \multicolumn{1}{c|}{0/1/1} &
  0\%/37.5\%/0\% &
  \multicolumn{1}{c|}{0/0/0} &
  37.5\%/37.5\%/0\% \\ \hline
\multicolumn{1}{|c|}{success rate} &
   &
  \multicolumn{1}{c|}{} &
  64.7\% &
  \multicolumn{1}{c|}{} &
  64.7\% &
  \multicolumn{1}{c|}{} &
  70.6\% &
  \multicolumn{1}{c|}{} &
  58.8\% &
  \multicolumn{1}{c|}{} &
  41.2\% &
  \multicolumn{1}{c|}{} &
  47.1\% \\ \hline
\multicolumn{1}{|c|}{\multirow{18}{*}{RTLLM\cite{lu2023rtllm}}} &
  accu &
  \multicolumn{1}{c|}{2} &
  0\% &
  \multicolumn{1}{c|}{5} &
  0\% &
  \multicolumn{1}{c|}{5} &
  0\% &
  \multicolumn{1}{c|}{0} &
  0\% &
  \multicolumn{1}{c|}{3} &
  0\% &
  \multicolumn{1}{c|}{5} &
  0\% \\ \cline{2-14} 
\multicolumn{1}{|c|}{} &
  adder\_8bit &
  \multicolumn{1}{c|}{4} &
  0\% &
  \multicolumn{1}{c|}{4} &
  7\% &
  \multicolumn{1}{c|}{0} &
  100\% &
  \multicolumn{1}{c|}{5} &
  0\% &
  \multicolumn{1}{c|}{3} &
  0\% &
  \multicolumn{1}{c|}{3} &
  46\% \\ \cline{2-14} 
\multicolumn{1}{|c|}{} &
  adder\_16bit &
  \multicolumn{1}{c|}{1} &
  55\% &
  \multicolumn{1}{c|}{0} &
  52\% &
  \multicolumn{1}{c|}{0} &
  70\% &
  \multicolumn{1}{c|}{3} &
  55\% &
  \multicolumn{1}{c|}{2} &
  50\% &
  \multicolumn{1}{c|}{5} &
  0\% \\ \cline{2-14} 
\multicolumn{1}{|c|}{} &
  adder\_32bit &
  \multicolumn{1}{c|}{3} &
  0\% &
  \multicolumn{1}{c|}{5} &
  0\% &
  \multicolumn{1}{c|}{0} &
  0\% &
  \multicolumn{1}{c|}{5} &
  0\% &
  \multicolumn{1}{c|}{0} &
  0\% &
  \multicolumn{1}{c|}{0} &
  0\% \\ \cline{2-14} 
\multicolumn{1}{|c|}{} &
  adder\_64bit &
  \multicolumn{1}{c|}{2} &
  0\% &
  \multicolumn{1}{c|}{5} &
  0\% &
  \multicolumn{1}{c|}{5} &
  0\% &
  \multicolumn{1}{c|}{5} &
  0\% &
  \multicolumn{1}{c|}{5} &
  0\% &
  \multicolumn{1}{c|}{0} &
  0\% \\ \cline{2-14} 
\multicolumn{1}{|c|}{} &
  multi_16bit &
  \multicolumn{1}{c|}{0} &
  0\% &
  \multicolumn{1}{c|}{5} &
  0\% &
  \multicolumn{1}{c|}{5} &
  0\% &
  \multicolumn{1}{c|}{2} &
  0\% &
  \multicolumn{1}{c|}{5} &
  0\% &
  \multicolumn{1}{c|}{2} &
  0\% \\ \cline{2-14} 
\multicolumn{1}{|c|}{} &
  Johnson\_Counter &
  \multicolumn{1}{c|}{1} &
  97\% &
  \multicolumn{1}{c|}{5} &
  0\% &
  \multicolumn{1}{c|}{5} &
  0\% &
  \multicolumn{1}{c|}{4} &
  97\% &
  \multicolumn{1}{c|}{5} &
  0\% &
  \multicolumn{1}{c|}{5} &
  0\% \\ \cline{2-14} 
\multicolumn{1}{|c|}{} &
  right\_shifter &
  \multicolumn{1}{c|}{0} &
  100\% &
  \multicolumn{1}{c|}{0} &
  0\% &
  \multicolumn{1}{c|}{0} &
  0\% &
  \multicolumn{1}{c|}{0} &
  100\% &
  \multicolumn{1}{c|}{5} &
  0\% &
  \multicolumn{1}{c|}{0} &
  0\% \\ \cline{2-14} 
\multicolumn{1}{|c|}{} &
  mux &
  \multicolumn{1}{c|}{1} &
  100\% &
  \multicolumn{1}{c|}{1} &
  0\% &
  \multicolumn{1}{c|}{0} &
  100\% &
  \multicolumn{1}{c|}{5} &
  0\% &
  \multicolumn{1}{c|}{5} &
  0\% &
  \multicolumn{1}{c|}{0} &
  0\% \\ \cline{2-14} 
\multicolumn{1}{|c|}{} &
  counter\_12 &
  \multicolumn{1}{c|}{4} &
  0\% &
  \multicolumn{1}{c|}{2} &
  97\% &
  \multicolumn{1}{c|}{0} &
  93\% &
  \multicolumn{1}{c|}{2} &
  93\% &
  \multicolumn{1}{c|}{5} &
  0\% &
  \multicolumn{1}{c|}{0} &
  50\% \\ \cline{2-14} 
\multicolumn{1}{|c|}{} &
  signal\_generator &
  \multicolumn{1}{c|}{5} &
  0\% &
  \multicolumn{1}{c|}{5} &
  0\% &
  \multicolumn{1}{c|}{0} &
  33.30\% &
  \multicolumn{1}{c|}{0} &
  0\% &
  \multicolumn{1}{c|}{5} &
  0\% &
  \multicolumn{1}{c|}{5} &
  0\% \\ \cline{2-14} 
\multicolumn{1}{|c|}{} &
  serial2parallel &
  \multicolumn{1}{c|}{3} &
  0\% &
  \multicolumn{1}{c|}{5} &
  0\% &
  \multicolumn{1}{c|}{5} &
  0\% &
  \multicolumn{1}{c|}{5} &
  0\% &
  \multicolumn{1}{c|}{0} &
  0\% &
  \multicolumn{1}{c|}{0} &
  0\% \\ \cline{2-14} 
\multicolumn{1}{|c|}{} &
  edge\_detect &
  \multicolumn{1}{c|}{0} &
  100\% &
  \multicolumn{1}{c|}{0} &
  98\% &
  \multicolumn{1}{c|}{0} &
  96\% &
  \multicolumn{1}{c|}{1} &
  0\% &
  \multicolumn{1}{c|}{5} &
  0\% &
  \multicolumn{1}{c|}{0} &
  100\% \\ \cline{2-14} 
\multicolumn{1}{|c|}{} &
  width\_8to16 &
  \multicolumn{1}{c|}{3} &
  30\% &
  \multicolumn{1}{c|}{2} &
  33\% &
  \multicolumn{1}{c|}{0} &
  33\% &
  \multicolumn{1}{c|}{5} &
  0\% &
  \multicolumn{1}{c|}{0} &
  33\% &
  \multicolumn{1}{c|}{5} &
  0\% \\ \cline{2-14} 
\multicolumn{1}{|c|}{} &
  calendar &
  \multicolumn{1}{c|}{2} &
  100\% &
  \multicolumn{1}{c|}{5} &
  0\% &
  \multicolumn{1}{c|}{0} &
  100\% &
  \multicolumn{1}{c|}{5} &
  0\% &
  \multicolumn{1}{c|}{0} &
  100\% &
  \multicolumn{1}{c|}{0} &
  0\% \\ \cline{2-14} 
\multicolumn{1}{|c|}{} &
  RAM &
  \multicolumn{1}{c|}{5} &
  0\% &
  \multicolumn{1}{c|}{5} &
  0\% &
  \multicolumn{1}{c|}{5} &
  0\% &
  \multicolumn{1}{c|}{3} &
  50\% &
  \multicolumn{1}{c|}{5} &
  0\% &
  \multicolumn{1}{c|}{5} &
  0\% \\ \cline{2-14} 
\multicolumn{1}{|c|}{} &
  alu &
  \multicolumn{1}{c|}{5} &
  0\% &
  \multicolumn{1}{c|}{5} &
  0\% &
  \multicolumn{1}{c|}{5} &
  0\% &
  \multicolumn{1}{c|}{5} &
  0\% &
  \multicolumn{1}{c|}{5} &
  0\% &
  \multicolumn{1}{c|}{5} &
  0\% \\ \cline{2-14} 
\multicolumn{1}{|c|}{} &
  pe &
  \multicolumn{1}{c|}{0} &
  100\% &
  \multicolumn{1}{c|}{0} &
  100\% &
  \multicolumn{1}{c|}{0} &
  100\% &
  \multicolumn{1}{c|}{4} &
  0\% &
  \multicolumn{1}{c|}{0} &
  0\% &
  \multicolumn{1}{c|}{0} &
  0\% \\ \hline
\multicolumn{1}{|c|}{success rate} &
   &
  \multicolumn{1}{c|}{} &
  27.8\% &
  \multicolumn{1}{c|}{} &
  5.6\% &
  \multicolumn{1}{c|}{} &
  22.2\% &
  \multicolumn{1}{c|}{} &
  5.6\% &
  \multicolumn{1}{c|}{} &
  5.6\% &
  \multicolumn{1}{c|}{} &
  5.6\%\\ \hline
\multicolumn{1}{|c|}{All success} &
   &
  \multicolumn{1}{c|}{} &
  45.7\% &
  \multicolumn{1}{c|}{} &
  34.3\% &
  \multicolumn{1}{c|}{} &
  45.7\% &
  \multicolumn{1}{c|}{} &
  31.4\% &
  \multicolumn{1}{c|}{} &
  22.9\% &
  \multicolumn{1}{c|}{} &
  25.7\%\\ \hline
\end{tabular}%
}
\end{table*}

\subsubsection{Verilog Generation  Comparasion} 
The results in Table \ref{tab:evalgeneration} indicate that on the Thakur et al. benchmark, our 13B model achieves an 11.8\% improvement in pass rate compared to Thakur et al.\cite{benchmarking} and a 5.9\% improvement compared to GPT-3.5. 
Additionally, our 13B model improves the pass rate from 25.7\% to 45.7\% compared to Llama 2-FT (General Aug) 13B, demonstrating the effectiveness of our approach. It shows that our data augmentation framework outperforms the general data generation method on both benchmarks.

\subsubsection{Ablation Study}
To assess the effectiveness of the proposed data augmentation framework, we conduct an ablation study on the Llama2-13B model. We separately finetune LLama 2-13B using a general data generation framework(only code completion) and our proposed data augmentation framework on the same base dataset. The results are shown in Fig. \ref{fig:ablationstudy}. The results indicate that utilizing only code completion framework exhibits flaws in terms of natural language descriptions for Verilog. Besides, in Tab. \ref{tab:evalgeneration}, the success rate of only code completion is 25.7\%, which is 45.7\% when adding alignment stage. Therefore, we conclude that program analysis alignment takes an important role in fine-tuning large language models for Verilog code generation.
\subsubsection{Verilog Repair Comparasion} The benchmark for the Verilog code repair task is derived from syntax-error code generated by the Large Language Model (LLM). As shown in Tab. \ref{tab:evalrepair}, The performance of our 13B model has been improved by 37.9\% compared to GPT-3.5 and by 62.1\% compared to the pre-trained Llama2-13B model.
\subsubsection{EDA Tool Script Generation Comparasion}The benchmark comprises five different levels of EDA script generation tasks, such as the 'core area' case, which represents the setting of the core area in the EDA script. The model takes natural language descriptions  as input and produces the corresponding EDA scripts as output.
As shown in Tab. \ref{tab:evalscript},
The results demonstrate that our 13B model can generate accurate EDA scripts based on Silicon Compiler with just one query, surpassing the performance of models like GPT-3.5 and Thakur et al.\cite{benchmarking}.


\section{Conclusions}
LLM-based chip design has demonstrated substantial promise for automating Verilog and EDA script generation. However, finetuning is currently constrained by the availability of training data. This paper proposed and evaluated a design-data augmentation framework aimed at enhancing the finetuning of LLMs in Verilog code generation domain. Experimental results revealed that the accuracy of Verilog generation surpasses that of the current state-of-the-art open-source Verilog generation model, increasing from 58.8\% to 70.6\% with the same benchmark and outperforms GPT-3.5 in Verilog repair and EDA Script Generation with only 13B weights.

\begin{acks}
This work was supported by the National Natural Science Foundation of China under NSFC.62090024, 62222411, 62025404, 92373206, 62202453
\end{acks}
  \def\bibfont{\scriptsize}
\bibliographystyle{ieeetr}
\bibliography{sample-base}

\end{document}